\documentclass[11pt, a4paper]{article}
 \usepackage{moriond,epsfig,amssymb, graphicx}

\def\be{\begin{equation}}
\def\ee{\end{equation}}
\def\bea{\begin{eqnarray}}
\def\eea{\end{eqnarray}}

\begin{document}
\title{Measurement of Transverse Spin Effects at COMPASS}
\author{Anselm Vossen for the COMPASS Collaboration}

\address{anselm.vossen@physik.uni-freiburg.de, Physikalisches Institut\\
Universit\"at Freiburg, 79104 Freiburg, Germany}
\maketitle\abstracts{ 
By measuring transverse single spin asymmetries one has access to the transversity distribution function $\Delta_T q(x)$
and the transverse momentum dependent Sivers function $q_0^T(x,\vec{k}_T)$.
New measurements from identified hadrons and hadron pairs,
produced in deep inelastic scattering of a transversely polarized 
$^6LiD$ target are presented. 
The data were taken in 2003 and 2004 by the COMPASS collaboration using the muon beam of the CERN SPS 
at 160 GeV/c, resulting in small asymmetries.
 } %
\noindent
{\small¥{\it Keywords}: transversity, transverse spin asymmetry, Collins asymmetry, Sivers asymmetry, COMPASS}
\section{Introduction}
\begin{figure}
\begin{center}
\includegraphics[scale=0.2]{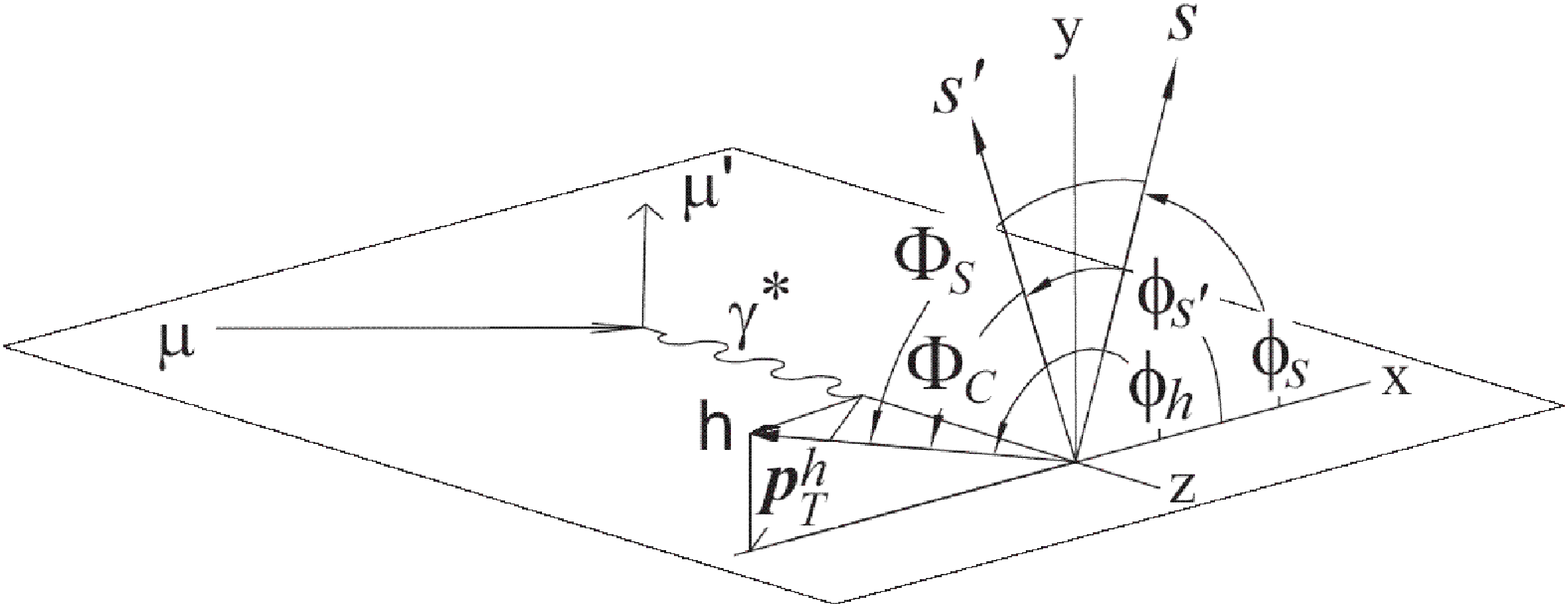}\includegraphics[scale=0.2]{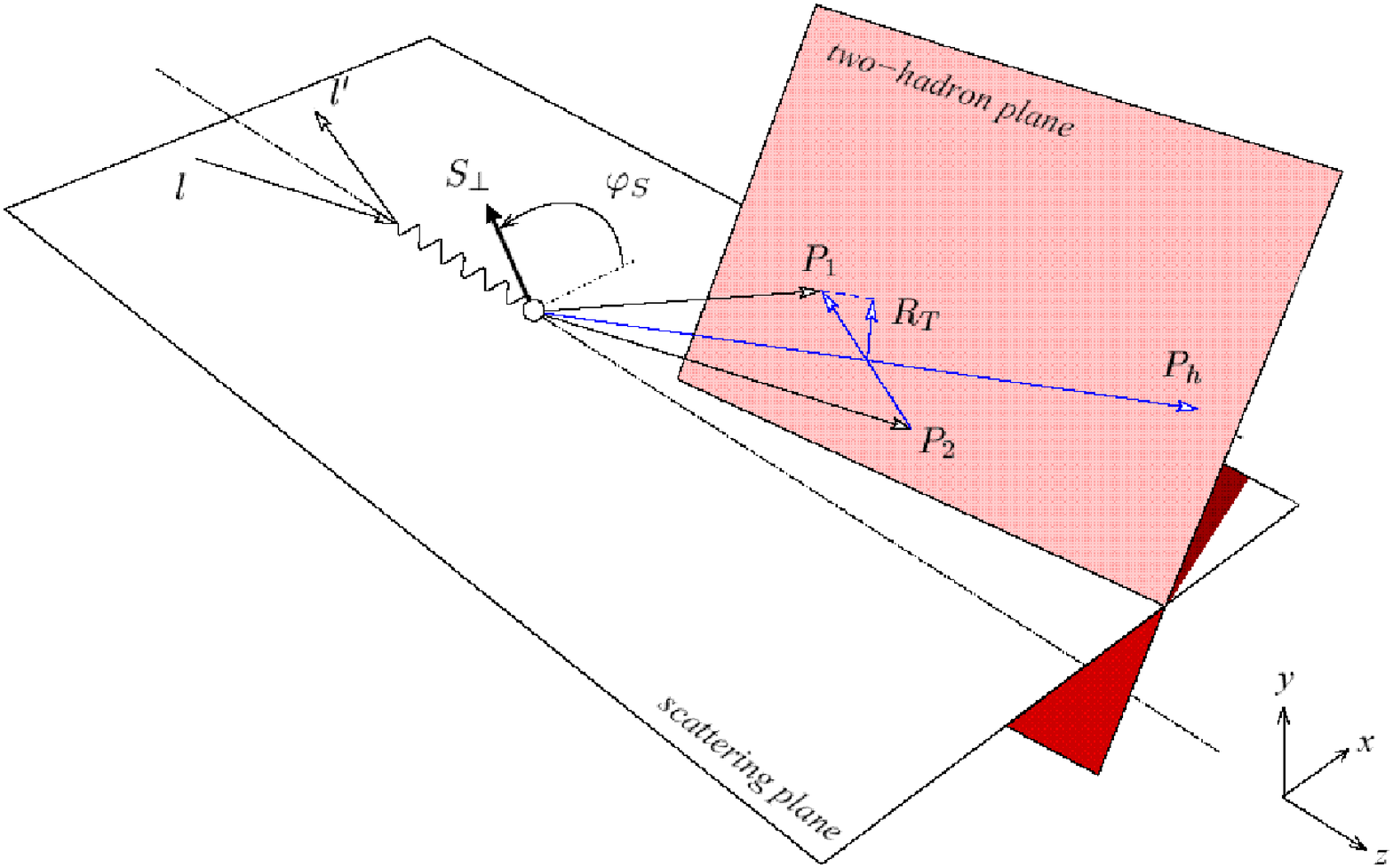}
\end{center}
\caption{Coordinate systems for azimuthal angles\label{cooSystems}. The x-z plane is defined by the incoming and scattered muon. The z-axis by the virtual photon
In the one hadron case (left) the initial state and final state polarization vectors are denoted s and s'. 
}
\end{figure}

Measurements \cite{fn}$^,$ \cite{an} showed, that the effects of transverse spin in high energy hadronic physics are 
not naturally suppressed, as it was assumed\cite{ka}.
On the contrary, transverse spin asymmetries provide a way for a measurement of transversity \cite{col}
and the quark transverse momentum $\vec{k}_T$ dependent distribution function $q_0^T(x,\vec{k}_T)$, 
the distribution of unpolarized quarks in a transversely polarized nucleon\cite{siv}.  
Denoting the Bjoerken scaling variable by $x$, the spin structure of the nucleon can be described at leading order
 by three leading twist distribution functions, the unpolarized quark distribution $q(x)$, 
the helicity distribution function $\Delta q(x)$, and the transversity $\Delta_T q(x)$. 
In the quark parton model (QPM) $\Delta_T q(x)$ can be interpreted as the difference of distributions  
of quarks polarized parallel and antiparallel to the nucleon spin in a transversely polarized nucleon.
Thus $\Delta q(x)$ and $\Delta_T q(x)$ are identical in a non-relativistic picture of the nucleon.
But in the QPM $\Delta q(x)$ can be thought of as describing the nucleon spin structure in a frame that is boosted parallel to the nucleon spin, whereas $\Delta_T q(x)$
in a frame that is boosted in the transverse direction. 
Since rotational symmetry is broken under boosts, $\Delta_T q(x)$ provides complementary information
to the proton spin puzzle.
The three distribution functions are of equal importance, however, in comparison to the first two, knowledge of the transversity is quite scarce,
since it remains inaccessible in inclusive DIS measurements due to its chiral odd nature.
However, semi inclusive measurements, where at least one hadron fragmenting via a chiral odd fragmentation function
 in the final state is detected, allow to probe transversity. 
Because the product of a distribution function and fragmentation function is again chiral even, it can be observed in transverse single spin
asymmetries.
The chiral odd fragmentation of a transversely polarized quark into a unpolarized hadron can be described by the Collins fragmentation function
$\Delta_T^0 D_q^h$ and the fragmentation into two unpolarized hadrons by the two hadron interference fragmentation function $H_1^\sphericalangle$.
Denoting the unpolarized fragmentation function by $D_q^h$, the spin dependent fragmentation of a quark can be written as
$D_q^h+\Delta_T^0 D_q^h \sin \Phi_C$ if one hadron is produced and $D_q^h + H_1^\sphericalangle \sin\Phi_R$\cite{ra}$^,$ \cite{jjt}$^,$ \cite{co}$^,$ \cite{ar} if two hadrons are produced.
$\Phi_C$, the Collins angle and $\Phi_R$ are the azimuthal angles between the polarization vector of the fragmenting quark 
and the momentum of the produced hadron and the vector $\vec{R}$, describing the two hadron system, respectively.
The relevant coordinate systems are depicted in fig. \ref{cooSystems}.
$\vec{R}$ is the linear combination of the momenta of the produced hadrons, weighted by the relative energy transfer from the scattered muon,
 to achieve a definition of azimuthal angles that is invariant against boosts along the photon direction: $\vec{R}=\frac{(z_2 \cdot \vec{P_1}-z_1 \cdot \vec{P_2})}{z_1+z_2}$.
By measuring the angular dependence of the produced hadrons on $\Phi_C$ and $\Phi_R$, respectively, it is possible to probe the transversity distribution.
For the Collins asymmetry one relies on the measurement of transverse momentum originating from the fragmentation process.
However, the azimuthal dependence on $\Phi_R$ of the cross section for the process $\Delta_T q \otimes H_1^\sphericalangle$ should remain after integrating out transverse momenta.
If one leaves the collinear picture and allows transverse momenta of the quarks, more distribution functions of quarks exist.
One of these, the so called Sivers function $q_0^T(x,\vec{k}_T)$, describes the distribution of unpolarized quarks in a transversely polarized nucleon.
It is strongly connected to the angular momenta of quarks in the nucleon, which might be another contribution to the nucleon spin.
The Sivers function can be probed via the Sivers effect, where the correlation of the quark transverse momentum with the nucleon spin leads
to the dependence of the SIDIS cross section on the azimuthal angle between the nucleon polarization vector and the momentum of the produced hadron.
This angle $\Phi_S$ is called Sivers angle.
Since the angular dependence of the cross section on $\Phi_S$ and $\Phi_C$ are orthogonal functions, Sivers and Collins effects can be disentangled  with a transversely polarized target.
\section{Experimental results}
\begin{figure}
\begin{center}
\includegraphics[scale=0.4]{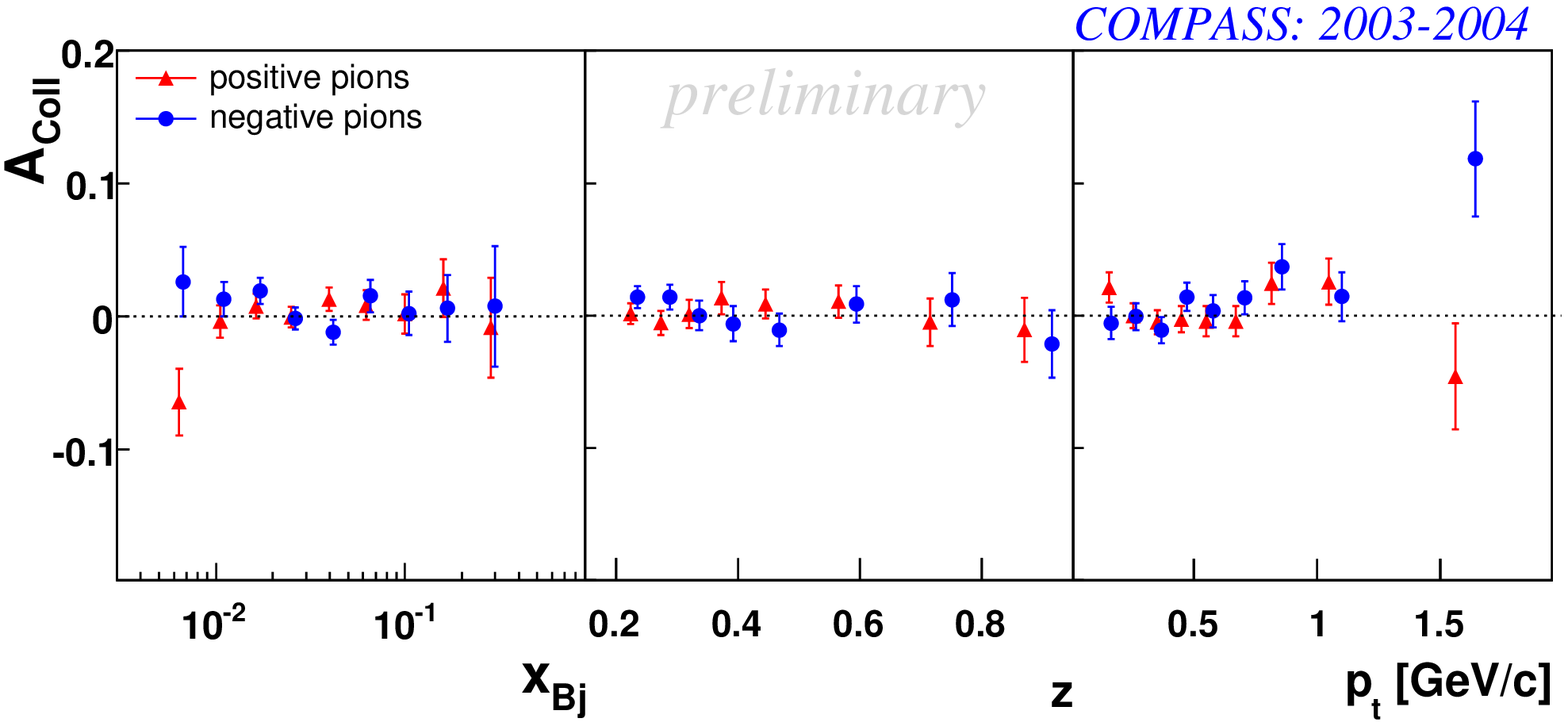}\includegraphics[scale=0.4]{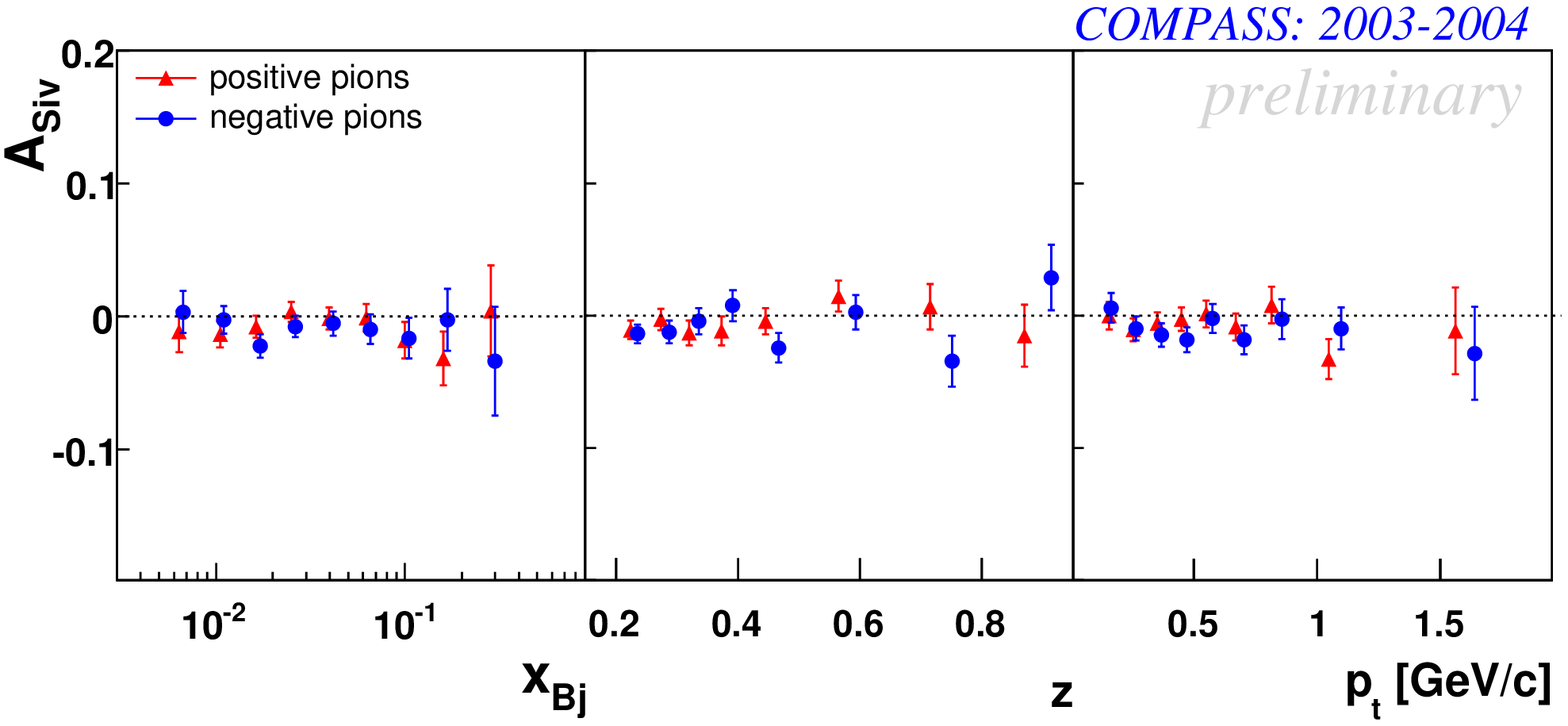}
\end{center}
\caption{Collins and Sivers asymmetries for pions\label{collSiversResults}}
\end{figure}

\begin{figure}
\centering
\includegraphics[scale=0.28]{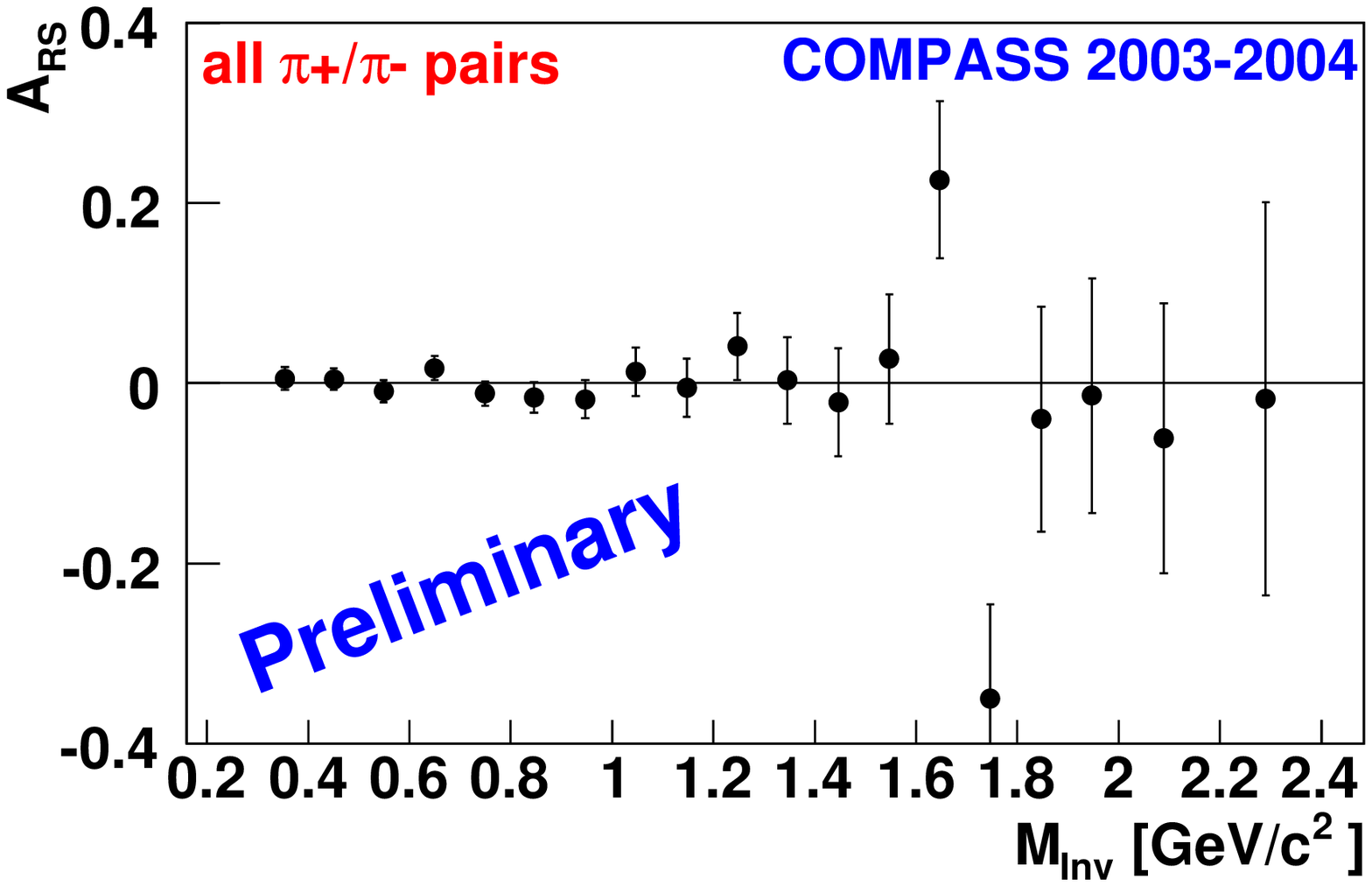}\includegraphics[scale=0.28]{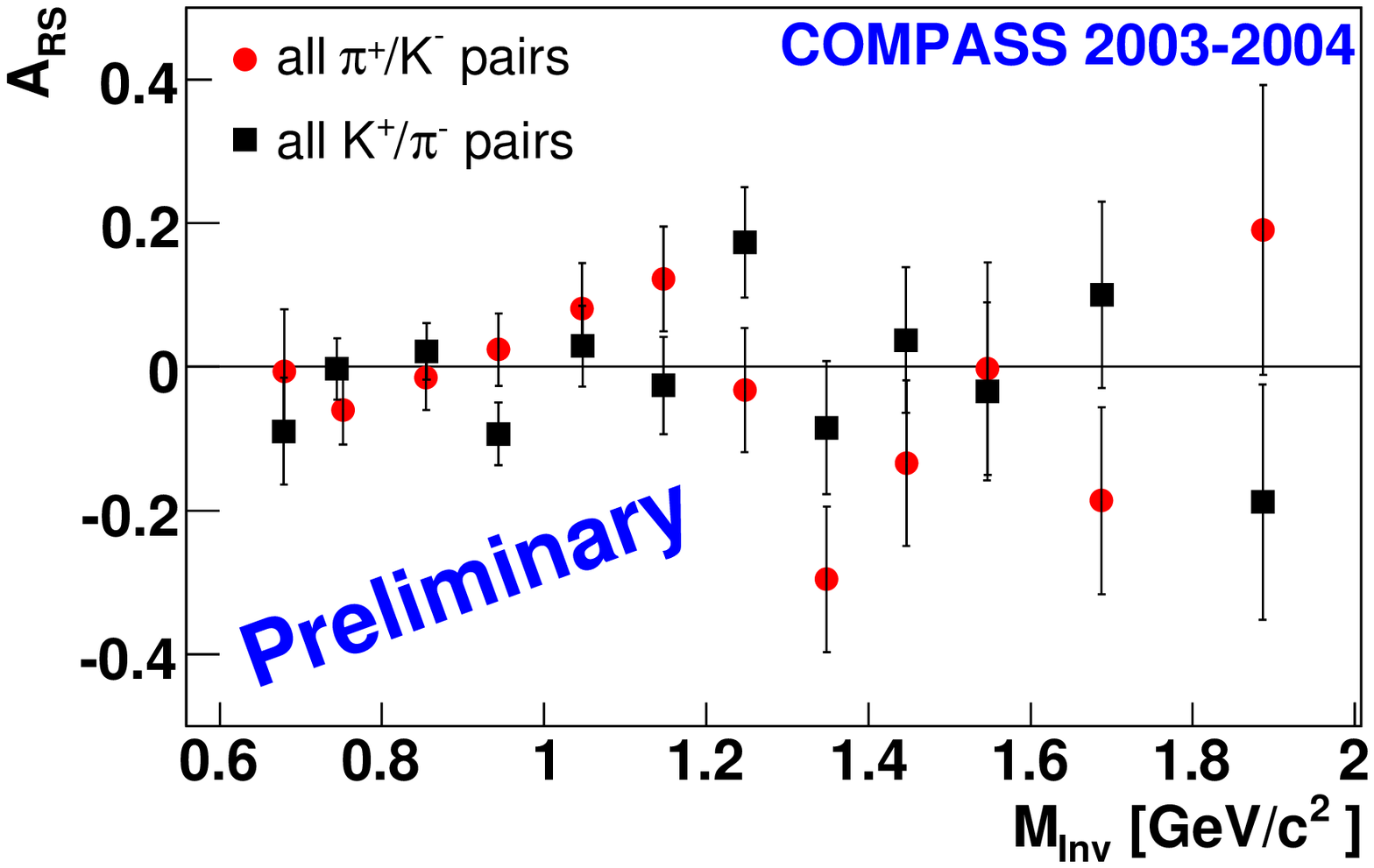}\includegraphics[scale=0.28]{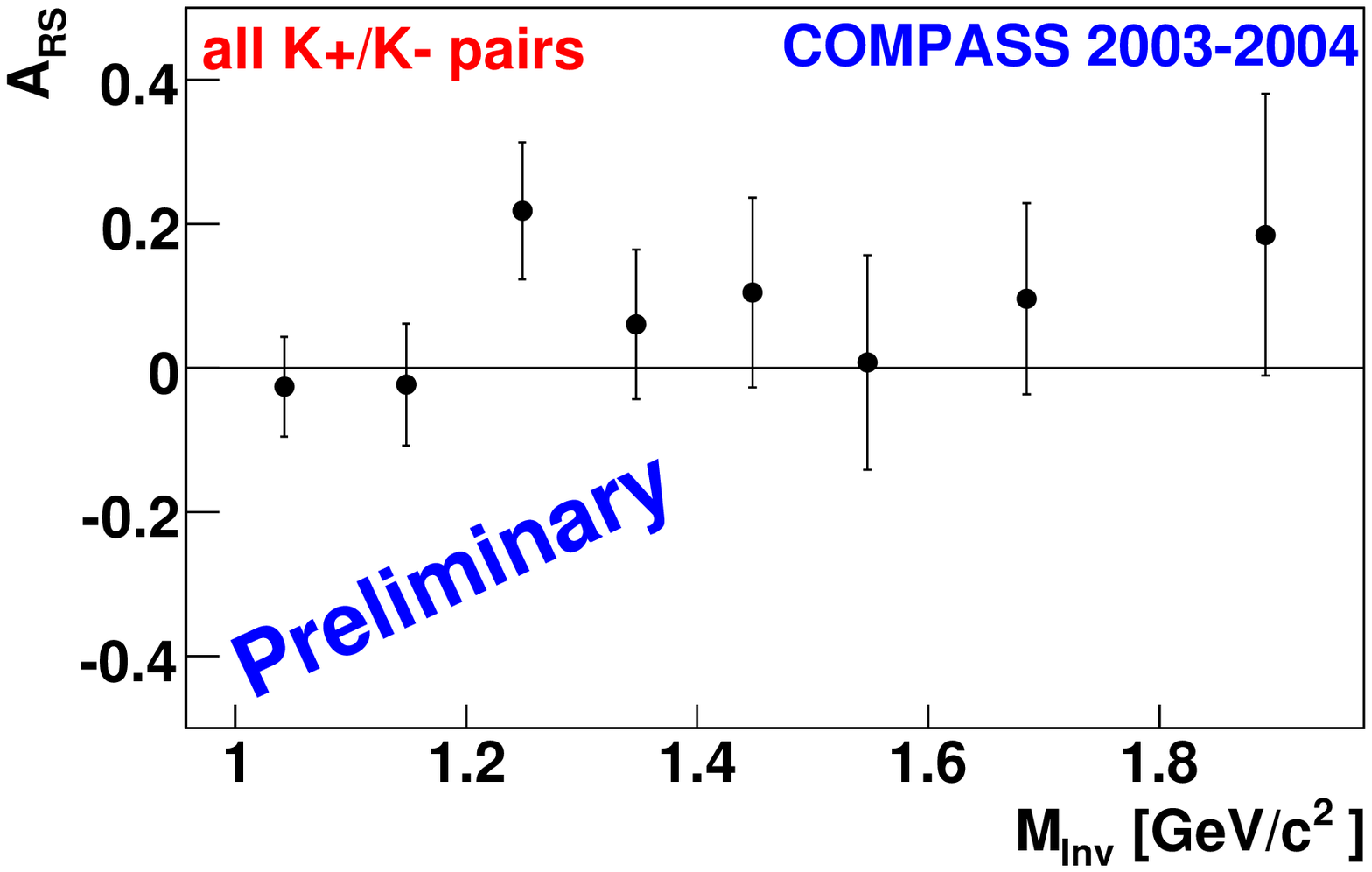}
\caption{Two hadron asymmetries in invariant mass binning. Due to space constraints not all binnings are shown.\label{twoHadronResults}}
\end{figure}

COMPASS is a fixed target experiment with a broad physics program at the M2 beamline at CERN and is described in detail in ref. \cite{sp}.
For about 20\% of the data taken in the years 2002, 2003 and 2004 a transversely polarized $^6LiD$ target is used, which has a favourable dilution factor of $f\simeq 0.4$
and a polarization of about 50\%..
The target consisted of two cells which were polarized in opposite directions and polarization was reversed every 4-5 days to minimize systematic effects.
A Ring Imaging Cherenkov (RICH-1) detector and two hadron calorimeters provide particle identification capabilities. RICH-1 \cite{ri} is a gas RICH with a 3m 
long $C_4F_{10}$ radiator. It is characterized by large transverse dimensions in order to cover the whole spectrometer acceptance  ($\pm 250 \times 200 \textrm{mrad}$)
and was operational during data taking in the years 2003 and 2004. Asymmetries for unidentified hadrons were already published \cite{tr}.
For the analysis presented, events with an incoming beam track crossing both target cells, a scattered muon track and at least
one outgoing hadron for Collins and Sivers asymmetry extraction or two outgoing hadrons with opposite charge for the two hadron correlation extraction, are selected.
Positive identification of the hadrons in the final data sample by RICH-1 was required.
Clean hadron and muon selection was achieved using the hadron calorimeters and considering the amount of traversed material.
To select DIS events, cuts on $Q^2>1 (GeV/c)^2$ and $W > 5 GeV/c^2$ were made, $Q^2$ being the photon virtuality and W the mass of the final hadronic state.  
Additional cuts are applied to ensure that from the hadron sample the relevant physics signal can be extracted. 
Requiring the relative energy in the muon scattering process $y$ to fulfill $0.1 < y < 0.9 $ limits the error due to radiative corrections (higher cut) but warrants that 
the energy loss of the scattered beam particle is high enough to allow for reliable tracking (lower cut).
For the one hadron asymmetries  a lower limit of 0.2 for the relative energy $z$ of the hadron is demanded.
The underlying reasoning is that in the string fragmentation process hadrons with a higher energy are more sensitive to the properties
of the struck quark spin.
For the two hadron correlation the cut is $z_1,z_2 >0.1$ and in addition $z_1+z_2<0.9$ to avoid the kinematic region of exclusive $\rho$ production.
After all the cuts, $5.3\cdot10^6$ positive pions, $4.6\cdot10^6$ negative pions and $9.5\cdot10^5$ positive kaons and $6.2\cdot10^5$ negative kaons remain for the single hadron analysis.
The two hadron correlation signal is extracted from  $3.7\cdot 10^6$ $\pi^+-\pi^-$, $2.4\cdot 10^5$ $\pi^+-K^-$, $3.0 \cdot10^5$ $K^+-\pi^-$ and $8.6\cdot 10^4$ $K^+-K^-$ pairs.
From these samples the respective asymmetries $A_j$ are extracted by azimuthal count rate asymmetries for target cells with different polarizations.
The count rate in the upstream and downstream target cell (k=u,d) for the two polarisations (+,$-$) $N^\pm_{j,k}$ in a given $\Phi_j$ bin (j=C,S,R) can be written
as 
\begin{equation}
N^\pm_{j,k}=F_k^\pm n_k \sigma a^\pm_{j,k} (\Phi_j) \cdot (1 \pm \epsilon_{j,k}^\pm \sin \Phi_j)
\end{equation}
Here F is the muon flux, n the number of target particles, $\sigma$ the spin averaged cross section, $a_j$ the product of the angular acceptance and efficiency 
of the spectrometer. The quantity $\epsilon_{j,k}^\pm $is $f\cdot |P^\pm_{T,k}|\cdot D_{NN}\cdot A_j$. Where $f$ is the dilution and $|P^\pm_{T,k}|$ the polarisation of the target, $D_{NN}$ the spin transfer
of the virtual photon to the fragmenting quark and $A_j$ the asymmetry\cite{tr}.
For Collins and two hadron asymmetries $D_{NN}$ can be calculated in QED as $D_{NN}=\frac{1-y}{1-y+y^2/2}$. Since the Sivers effect probes unpolarized quarks $D_{NN}$ 
does not enter into the corresponding asymmetry.
With the obtained count rates the double ratio product
\begin{equation}
A_j(\Phi_j)=\frac{N_{j,u}^+(\Phi_j)}{N_{j,u}^-(\Phi_j)}\cdot\frac{N_{j,d}^+(\Phi_j)}{N_{j,d}^-(\Phi_j)}
\end{equation}
is build and fitted with $p_0\cdot (1+A_j^m \sin \Phi_j)$. 
The raw asymmetry $A^m_j=\epsilon^+_{j,u}+\epsilon^+_{j,d}+\epsilon^-_{j,u}+\epsilon^-_{j,d}=4<\epsilon_j>$
 is calculated for kinematical binning in 
$z$, the relative energy of the produced hadron, $p_t$, its transverse momentum and $x$, the Bjoerken scaling variable in the one hadron
case, whereas it is build for bin in $z$, $M_{\mathrm{Inv}}$ and $x$ in the two hadron case. $z$ and $M_{Inv}$ denote the relative energy and invariant mass
of the two hadron system. 
The corrected  asymmetries are shown in figures \ref{collSiversResults} and \ref{twoHadronResults}.
Phenomenological work on the Sivers effect\cite{ans} has shown that HERMES results for protons and COMPASS results on 
deuteron may be described within the the same theoretical frame, at least at the present level of accuracy of the data.
The obtained asymmetries are small and agree well with model calculations, that predict suppressed signals due to the isoscalar target.
A model of transversity from the chiral quark soliton model and Collins fragmentation function extracted from 
a fit to HERMES proton data \cite{egs} shows that the favoured and unfavoured Collins fragmentation function seem to be of the same magnitude but of opposite sign. 
The predictions obtained from this model agree well with the measured asymmetries for unidentified hadrons.
Similarly the predictions for the two hadron asymmetries depending on the convolution of transversity with $H_1^\sphericalangle$ are predicted to be small \cite{ra}. 
Due to an interference term in the two pion production one model \cite{jjt} predicts a strong dependence on the invariant mass around the $\rho$ mass, which 
cannot be observed in the current COMPASS data.
Measurements taken so far on a deuteron target allow constraints on models for the d-quark Sivers and transversity distribution\cite{tr}. 
They also point to the absence of a gluon contribution to the orbital angular momentum of the partons in the nucleon. \cite{bg}
\section{Outlook} 
COMPASS continues data taking in 2007 with a transversely polarized proton target. In combination with the data already measured on deuteron 
flavour separation for transversity and Sivers distribution function will be possible. An additional analysis is planned for 
leading hadron pair asymmetries
where according to \cite{co}$^,$ \cite{ar} an enhancement of the signal might be seen.

\end{document}